\documentclass[12pt]{amsart}

\usepackage{color}
\usepackage{algpseudocode}
\usepackage{algorithm}

\usepackage{mathtools}

\usepackage[pdftex]{graphicx}
\usepackage{epstopdf}
\graphicspath{{./fig/}{./fig/gluala/}{./fig/water/}}

\usepackage{caption}
\usepackage{subcaption}
\usepackage{amsmath}
\usepackage{listings}
\usepackage{hyperref} 
\usepackage{amssymb}
 
 \usepackage{type1cm}
 \usepackage[latin1]{inputenc}
 \usepackage[T1]{fontenc}
 \usepackage{mathptmx}

\usepackage{enumitem}

\newcommand{\OO}{\mathcal{O}}
\newcommand{\tr}[1]{\ensuremath{\text{trace}\left(#1\right)}} 
\newcommand{\CPPcode}[1]{\lstinline{#1}}

\newcommand{\norm}[1]{\ensuremath{\left\lVert#1\right\rVert}}
\newcommand{\wt}[1]{\ensuremath{\widetilde#1}}
\newcommand{\Xc}{\mathcal{X}}
\newcommand{\nocc}{\ensuremath{n_\text{occ}}}

\newcommand{\ie}{i.e.\ }

\title[Density matrix on distributed systems]{Efficient computation of the density matrix with \\ error control on distributed computer systems}
\author[A. Kruchinina, E. Rudberg and E. H. Rubensson]{Anastasia Kruchinina, Elias Rudberg and Emanuel H. Rubensson}
\date{\today}
\begin{document}
\begin{abstract}
  The recursive polynomial expansion for construction of a density matrix approximation with rigorous error control [J. Chem. Phys. 128, 074106 (2008)] is implemented in the quantum chemistry program {\sc  Ergo} [SoftwareX 7, 107 (2018)] using the Chunks and Tasks matrix library [Parallel Comput. 57, 87 (2016)]. The expansion is based on second-order polynomials and accelerated by the scale-and-fold technique [J. Chem. Theory Comput. 7, 1233 (2011)]. We evaluate the performance of the implementation by computing the density matrix from the Fock matrix in the large-scale self-consistent field calculations. 
  We demonstrate that the amount of communicated data per worker process tends to a constant with increasing system size and number of computer nodes such that the amount of work per worker process is fixed.
\end{abstract}
\maketitle

\section{Introduction}

The density matrix $D$ is constructed from the Fock/Kohn-Sham matrix $F$ in each step of self-consistent field (SCF) calculations emerging in the Hartree--Fock method~\cite{Roothaan} and Kohn--Sham density functional theory~\cite{hohen,KohnSham65}.
In orthogonal basis set, the matrix $F$ has eigenvalues $\lambda_i$ and corresponding eigenvectors $y_i$:
\begin{align}
Fy_i = \lambda_i y_i.
\end{align}
Each eigenpair (eigenvalue and eigenvector) corresponds to one molecular orbital. 
Let the eigenvalues of the matrix $F$ be arranged
in ascending order
\begin{align}
 \lambda_{1} \leq \lambda_{2} \leq \ldots \leq \lambda_{\textrm{homo}}
 <
\lambda_{\textrm{lumo}} \leq \ldots \leq \lambda_{N-1} \leq \lambda_{N},
\end{align}
where $\lambda_{\textrm{homo}}$ is the eigenvalue corresponding to the
highest occupied molecular orbital (homo eigenvalue), and $\lambda_{\textrm{lumo}}$ is the eigenvalue corresponding to the lowest unoccupied molecular orbital (lumo eigenvalue). We assume that the homo-lumo gap $\xi := \lambda_{\textrm{lumo}} - \lambda_{\textrm{homo}}$ is  non-zero. The number of occupied orbitals we denote with $\nocc$.
The subspace spanned by the eigenvectors of the matrix $F$, that correspond to the occupied molecular orbitals, is referred to as the \textit{occupied
  invariant subspace} of $F$.
At zero temperature the density matrix is the matrix for orthogonal
  projection onto the occupied invariant subspace of $F$:
\begin{align}
D = \sum_{i=1}^{\nocc} y_iy_i^T.
\label{eq:eig_problem_D}
\end{align}
Computation of the density matrix using~\eqref{eq:eig_problem_D}  requires diagonalization of the matrix $F$ which scales cubically with the matrix size, limiting the calculations to rather small systems. Alternative density matrix methods see the density matrix construction as an evaluation of a matrix function
\begin{align}
 D &= \theta(\mu I - F),\\
\theta(x) &= 
\begin{cases}
0 & \text{ if } x < 0,\\
1 & \text{ otherwise},
\end{cases}
\end{align}
where $\mu$ is located in the homo-lumo gap. 
Recursive polynomial expansions, also known as purification methods, utilize low-order polynomials $f_i$, $i=0,1,2,\ldots$ to approximate the shifted and scaled step function $\theta(\mu-x)$:
\begin{align}
 D \approx X_n = f_n(f_{n-1}(\ldots f_0(F)\ldots)).
\end{align}
Recursive polynomial expansions are implemented in a number of electronic structure codes including {\sc CP2K}~\cite{VandeVondele_2012}, {\sc  Ergo}~\cite{ergo_web,ErgoSoftwareX}, {\sc  Honpas}~\cite{honpas}, and {\sc LATTE}~\cite{LATTE-jcp-2012}. 
The initial polynomial $f_0$ usually maps the eigenvalues of $F$ into the interval $[0,\ 1]$ in reverse order.
The choice of low-order polynomials $f_i$, $i\geq1$ depends on the   method, but typically one or two matrix multiplications per iteration are required. Niklasson~\cite{Nikl2002} proposed an efficient recursive expansion algorithm based on the second-order spectral projection polynomials $x^2$ and $2x-x^2$. We will refer to this algorithm as the SP2 algorithm. In the original SP2 algorithm, the polynomial in each iteration is chosen based on the matrix trace, which may result in slow convergence in some cases. The convergence can be improved by selecting polynomials based on the homo and lumo eigenvalues~\cite{ErrorControl}. This will also remove the explicit dependence on the system size. In the scheme proposed in~\cite{interior_eigenvalues_2014}, homo and lumo eigenvalue estimates are obtained at negligible cost using Frobenius norms and traces of intermediate matrices $X_i-X^2_i$. The homo and lumo estimates for the current matrix $F$ are obtained at the end of the recursive expansion, but they can be propagated to the following SCF cycle and used there in the recursive expansion. In the first SCF cycles, where the eigenvalue bounds are not available yet or not accurate enough, we use the trace-correcting recursive expansion mentioned above~\cite{Nikl2002,ErrorControl}. Knowledge of homo and lumo eigenvalue estimates allows for further improvement of the SP2 scheme convergence by making use of the scale-and-fold technique proposed in~\cite{Rub2011}. The SP2 scheme, accelerated using the scale-and-fold technique, we refer to as the SP2ACC recursive expansion. We use the parameterless stopping criteria for recursive expansions developed in~\cite{stop_crit_2016}, which automatically stop the iterations when the level of numerical errors is reached and the final density matrix approximation cannot be substantially improved.

Matrices, obtained in electronic structure calculations,  posses the exponential decay property of matrix element magnitude~\cite{benzi_decay,localized_inverse_factorization}. After removal of small matrix elements (truncation), matrices are sparse, and the matrix sparsity can be preserved throughout the recursive expansion iterations. Another consequence of the exponential decay property is that non-negligible matrix elements tend to be localized in certain regions. We implement the truncation scheme proposed in~\cite{ErrorControl}, allowing for rigorous control of the error in the occupied invariant subspace, which is a measure of the density matrix approximation error. The error is controlled in the spectral norm. However, we use a computationally cheaper upper bound of the spectral norm for selecting truncation threshold values in each iteration. For that purpose, we propose the Frobenius-infinity mixed matrix norm, which provides an upper bound of the spectral norm for Hermitian matrices, and which has the same asymptotic behavior as the spectral norm for matrices with localization properties of element magnitudes.

The main computational cost in the recursive expansion comes from sparse matrix multiplications, which makes it attractive for parallel computations. Parallel algorithms, developed for dense matrix-matrix multiplication, take advantage of a uniform distribution of matrix elements over computational resources. A few efficient algorithms for sparse matrix-matrix multiplication use similar ideas. A uniform distribution of non-zero matrix elements can be obtained dynamically during runtime by a random permutation of matrix rows and columns (see e.~g.~\cite{azad2016exploiting,bulucc2012parallel,dawson2018massively,lazzaro2017increasing}). The permuted matrices are divided into blocks with a sparse matrix representation of elements at the block level. Then, some algorithm, inspired by the parallel dense matrix-matrix multiplication, is used for managing the communication between computer nodes. Individual blocks are multiplied locally using a chosen algorithm for sparse matrix-matrix multiplication. The communication cost per node of the 2D algorithm, utilizing a two-dimensional process grid~\cite{azad2016exploiting}, scales as $\OO(\sqrt{N})$ with the matrix size $N$ in a weak scaling regime. The parallel performance can be improved by introducing a third process grid dimension~\cite{azad2016exploiting,dawson2018massively,lazzaro2017increasing}. Compared to the 2D approach, in a weak scaling limit the communication cost per node of the 3D algorithm is constant~\cite{ballard2013communication}.
Methods involving random permutation were originally developed for the general sparse matrix-matrix multiplication.  By randomly permuting matrix rows and columns, any locality information is lost. Furthermore, permutations of matrix rows and columns may be needed for some intermediate matrices in the recursive expansion to enforce load balance. An efficient sparse matrix-matrix multiplication, utilizing the locality of non-zero elements, has been implemented in the Chunks and Tasks matrix library (CHTML)~\cite{LocalityAwareRubensson2016}. It has been shown that the dynamic work and data distribution over computational resources allows for efficient utilization of data locality and constant communication cost per worker process in a weak scaling regime~\cite{LocalityAwareRubensson2016}.

In this work, we expand CHTML with matrix operations required in the recursive expansion, and implement the whole recursive expansion using CHTML in the quantum chemistry code Ergo~\cite{ErgoSoftwareX}. We evaluate the weak scaling performance of our implementation using the Frobenius-infinity mixed norm for the estimation of the spectral norm in the error control scheme. We request the same accuracy of the density matrix approximation for all system sizes.
The usage of the Frobenius-infinity mixed norm allows us to keep the number of non-zero elements per row asymptotically constant with increasing system size. We show numerically that the communication cost per node is constant in a weak scaling limit.

\section{Error control}

A common approach to preserve matrix sparsity is to remove blocks of elements with norms smaller than some predefined threshold value. Often such threshold values are provided by the user and not explicitly related to the desired accuracy of the density matrix approximation.
 In~\cite{ErrorControl}, an upper bound of the forward error $\norm{D-X_n}$ is given by
\begin{align}
  \norm{D-X_n} \leq  \norm{D-P_{\Xc_n}} + \norm{P_{\Xc_n}-X_n},
  \label{eq:upper_bound_forw_err}
\end{align}
where \norm{\cdot} is any unitary-invariant matrix norm, $P_{\Xc_n}$ is the matrix for orthogonal projection onto the occupied subspace $\Xc_n$ of the matrix $X_n$.
The first norm $\norm{D-P_{\Xc_n}}$ measures erroneous rotations of the occupied invariant subspace or errors in eigenvectors. A scheme for keeping the error in the occupied invariant subspace smaller than a desired accuracy $\gamma$ is proposed in~\cite{ErrorControl}. Let $\widetilde{X}_i = X_i+E_i$ be the matrix obtained after truncation of $X_i$,
where the truncation is written as an explicit perturbation of $X_i$ by the error matrix $E_i$. Then, it has been shown in~\cite{ErrorControl}, that
\begin{align}
  \norm{D-P_{\Xc_n}} \leq \sum_{i=0}^{n} \frac{\norm{E_i}}{\tilde{\xi}_i - \norm{E_i}_2},
\end{align}
where $\tilde{\xi}_i$ is a lower bound of the homo-lumo gap in iteration  $i$ of the recursive expansion. Assume that an upper bound $n_{\text{max}}$ of the total number of iterations in the recursive expansion is known. Then, if the truncation threshold values $\tau_i$ are chosen such that:
\begin{align}
  \tau_i := \frac{\frac{\gamma}{n_{\text{max}}+1}\tilde{\xi}_i}{1+\frac{\gamma}{n_{\text{max}}+1}},
  \label{eq:tau_trunc}
\end{align}
and the truncation is performed so that 
\begin{align}
  \norm{E_i} \leq \tau_i,
  \label{eq:error_tau_trunc}
\end{align}
then the total error in the occupied subspace is bounded by $\gamma$:
\begin{align}
  \norm{D-P_{\Xc_n}} \leq \gamma.
\end{align}
The error in the occupied subspace is increasing with the number of iterations, and any lost accuracy cannot be recovered in later iterations of the recursive expansion.

As mentioned above, any unitary-invariant norm can be used in the error control scheme. Examples of unitary-invariant norms are the spectral and Frobenius norms.  
The spectral $\norm{\cdot}_2$ and Frobenius $\norm{\cdot}_F$ norms of a Hermitian matrix $A = (a_{ij})$ of size $N$ are defined as:
\begin{align}
    \norm{A}_2 &= \rho(A),\\
    \norm{A}_F &= \sqrt{\sum_{i,j=0}^{N}a_{ij}^2},
\end{align}
where the spectral radius $\rho(A)$ is equal to the largest by absolute value eigenvalue of $A$.
The spectral norm depends only on the eigenvalue distribution, but it is expensive to compute for large systems. However, the computationally cheaper Frobenius norm depends on the system size, making its usage inappropriate in large scale calculations. Thus, we use the spectral norm in the error control scheme.  If we can measure some upper bound $\hat{S}_i$ to the spectral norm of $E_i$, inequality~\eqref{eq:error_tau_trunc} can be satisfied by making sure that 
$\hat{S}_i \leq \tau_i$.

In the presence of numerical errors the computed density matrix approximation slightly deviates from idempotency. The second norm $\norm{P_{\Xc_n}-X_n}$ in~\eqref{eq:upper_bound_forw_err} measures errors in the density matrix approximation coming from errors in eigenvalues. 
In~\cite{stop_crit_2016}, we proposed a parameterless stopping criteria for the recursive expansions, which do not require any user defined tolerance. The criteria are satisfied when the numerical errors in the density matrix approximation, coming from the removal of small matrix elements and round-off errors, prevent any further improvement of the result. The idea is to observe the numerically computed convergence order during the recursive expansion and automatically detect the convergence order drop. The convergence order in iteration $i$ is  controlled using  the idempotency error $\norm{X_i-X^2_i}$, which can be considered as a measure of the error $\norm{P_{\Xc_i}-X_i}$. Note that the error $\norm{P_{\Xc_n}-X_n}$ may be decreased to a required level of accuracy by reducing truncation in the last recursive expansion iterations. However, due to the increased computational cost, we do not use this possibility here, and choose truncation values given by~\eqref{eq:tau_trunc} throughout the whole recursive expansion. If the SP2ACC scheme is used, we fall back to the SP2 scheme in the final iterations when acceleration does not have a strong effect on the convergence, and use the stopping criterion developed for the SP2 scheme, as suggested in~\cite{stop_crit_2016}. In~\cite{stop_crit_2016}, the proposed stopping criteria use the spectral norm. In addition, it is shown numerically, that the Frobenius norm can be used instead of the spectral norm. In contrast to the error control scheme above, the dependence of the Frobenius norm on the system size is less critical in the stopping criterion, since the Frobenius norm provides a good estimate to the spectral norm close to convergence.

\begin{sloppypar}
In the following section we consider matrix norms providing upper bounds to the spectral norm. Such norms can be used to ensure~\eqref{eq:error_tau_trunc}.
\end{sloppypar}

\section{Upper bounds for the spectral norm}

In~\cite{mixedNormTrunc}, the Frobenius-spectral mixed norm is proposed. Let a matrix $A$ be partitioned into submatrices:
\begin{align}
  A = 
  \begin{pmatrix}
  A_{11}  & A_{12} \\
  A_{21}  & A_{22}
\end{pmatrix}.
\label{eq:sumb_of_A}
\end{align}
Then the Frobenius-spectral mixed norm of the matrix $A$ is defined as:
\begin{align}
  \norm{A}_{m} = 
  \norm{
  \begin{pmatrix}
    \|A_{11}\|_F  & \|A_{12}\|_F \\
    \|A_{21}\|_F  & \|A_{22}\|_F
  \end{pmatrix}}_2.
\end{align}
The Frobenius-spectral mixed norm gives an upper bound of the spectral norm~\cite{mixedNormTrunc}:
\begin{align}
  \norm{A}_{2} \leq \norm{A}_{m}.
\end{align}
Moreover, the Frobenius-spectral mixed norm has the same asymptotic behavior as the spectral norm, and is computationally cheaper than the spectral norm~\cite{mixedNormTrunc}.  

In this work, we introduce a Frobenius-infinity mixed norm, which we refer to as Frob-Inf and denote $\norm{\cdot}_M$, obtained by the combination of the Frobenius and infinity matrix norms.  The infinity norm of a matrix $A = (a_{ij})$ of size $N$ is equal to the largest sum of absolute values of elements in each matrix row:
\begin{align}
  \norm{A}_\infty = \max_{1\leq i \leq N}{\sum_{j=1}^{N}|a_{ij}|}.
\end{align}
If the matrix $A$ is divided into submatrices as in~\eqref{eq:sumb_of_A}, the Frob-Inf norm of the matrix $A$ is defined as:
\begin{align}
  \norm{A}_{M} = 
  \norm{
  \begin{pmatrix}
    \|A_{11}\|_F  & \|A_{12}\|_F \\
    \|A_{21}\|_F  & \|A_{22}\|_F
  \end{pmatrix}}_\infty.
\end{align}
In practice, with increasing matrix size the submatrix size is kept fixed, and the number of submatrices in Frobenius-spectral and Frob-Inf norms is increasing. Thus, the Frobenius-spectral mixed norm requires computation of the spectral norm of matrices of growing size, which may be expensive for very large systems. In addition, efficient implementation of the spectral norm is difficult on distributed systems.

Under the assumption that the matrix $A$ is Hermitian, the infinity norm gives an upper bound of the spectral norm~\cite{horn2012matrix}. Thus, we have
\begin{align}
  \norm{A}_{2} \leq \norm{A}_{m} \leq \norm{A}_{M},
\end{align}
which proves that the Frob-Inf norm is an upper bound of the spectral norm. We note that the infinity norm can be used for estimating the spectral norm in the error control scheme. However, due to the simplicity of the implementation on distributed systems we chose to use the Frob-Inf norm instead.

\begin{figure}
  \begin{subfigure}[b]{.49\linewidth}
    \centering \includegraphics[width=\textwidth]{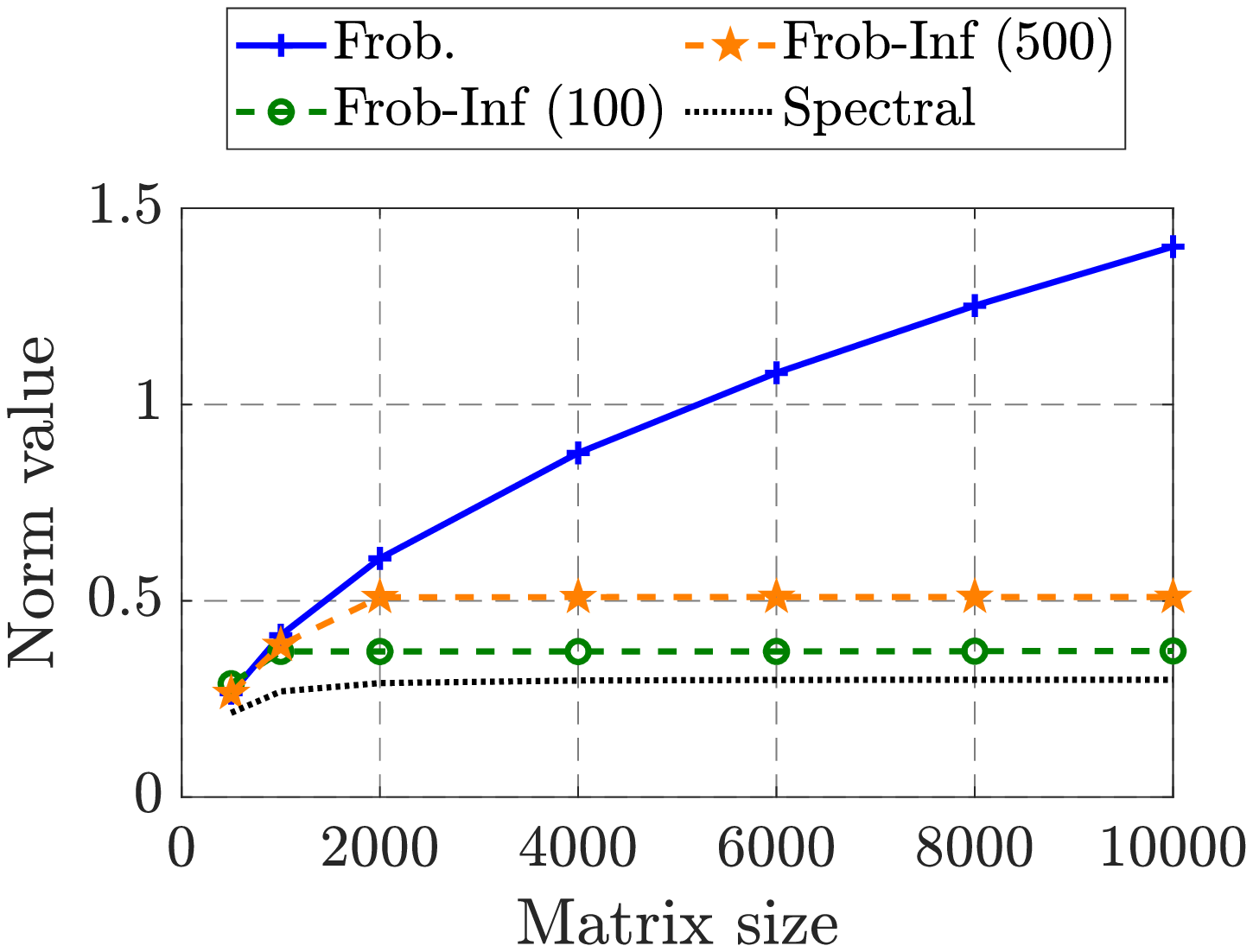}
    \caption{Banded matrices}\label{fig:matrix_norms_band}
  \end{subfigure}
  \begin{subfigure}[b]{.49\linewidth}
    \centering \includegraphics[width=\textwidth]{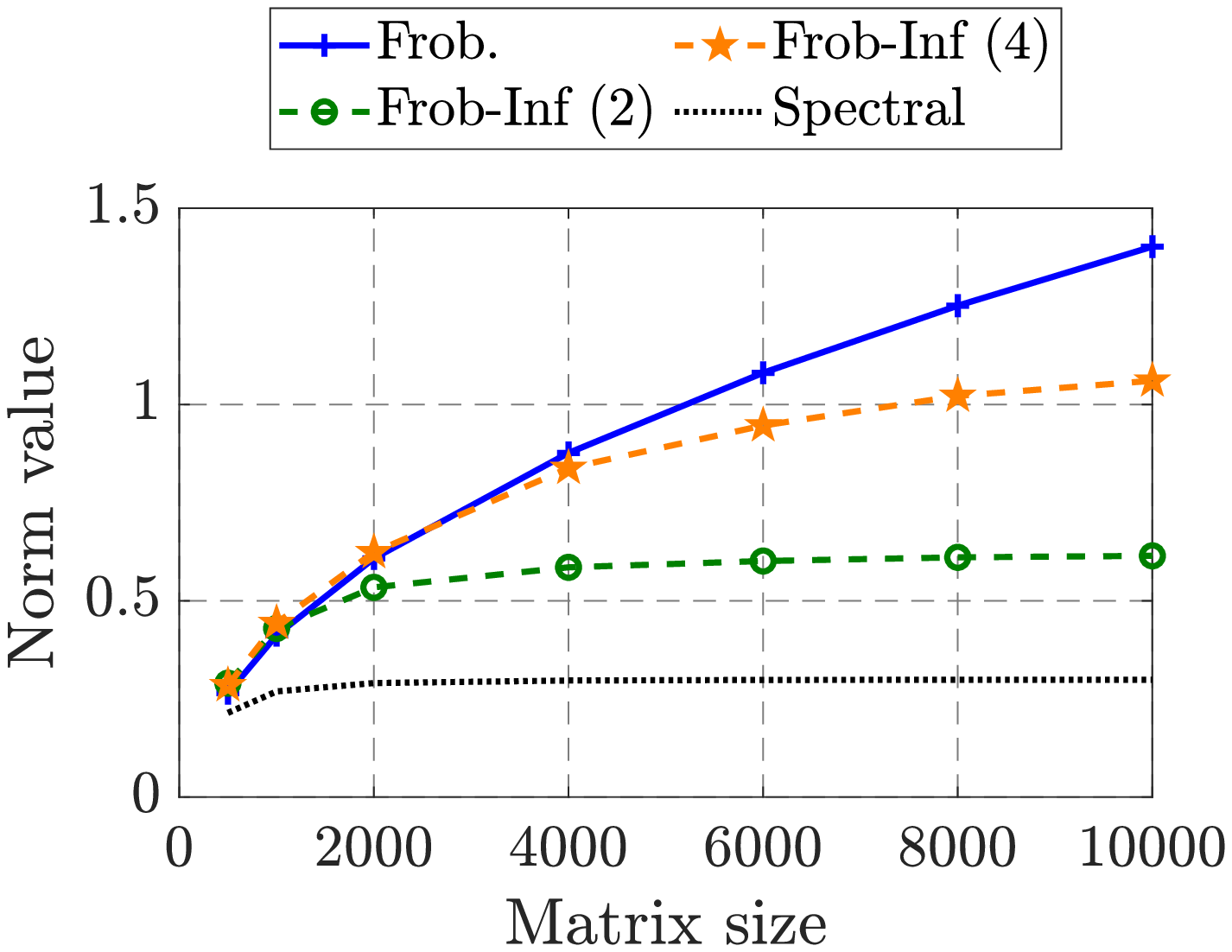}
    \caption{Randomly permuted matrices}\label{fig:matrix_norms_random}
  \end{subfigure}
  \caption{Comparison of spectral, Frobenius (``Frob'') and Frob-Inf norms. The number given in parenthesis for the Frob-Inf norm is the used block size. In the left panel we present results for banded matrices with bandwidth $2\times 300 + 1$. In the right panel we perform the random permutation of rows and columns of the banded matrices shown in the left panel. 
  }
  \label{fig:matrix_norms}
\end{figure}

We perform numerical tests in Matlab and compare values of matrix norms. In Figure~\ref{fig:matrix_norms_band} we show results for a symmetric banded matrix with random elements drawn from the standard uniform distribution on the open interval $]0, \, 10^{-3}[$ and bandwidth $2\times 300 + 1$.
The Frob-Inf norm, computed with block sizes 100 and 500, is much smaller than the Frobenius norm and has the same asymptotic behavior as the spectral norm. However, symmetric random permutation of rows and columns delays the moment when the Frob-Inf norm becomes proportional to the spectral norm compared to the non-permuted case. Note that in Figure~\ref{fig:matrix_norms_random} we use much smaller block sizes than in Figure~\ref{fig:matrix_norms_band}. 
It is hard to predict the behavior of the Frob-Inf norm for general matrices. However, if matrices posses the locality property and number of non-zeros per row does not grow significantly with matrix size, we can expect that, for large enough matrix sizes, the Frob-Inf norm will be proportional to the spectral norm, and our practical experiments support it.

\section{Density matrix construction}
\label{sec:dens_constr}

In this section we describe the density matrix construction using the SP2ACC recursive expansion implemented in the quantum chemistry program Ergo~\cite{ErgoSoftwareX} and parallelized using CHTML. Assuming that homo and lumo estimates are available, the density matrix algorithm based on the SP2ACC expansion may be divided into three steps:
\begin{enumerate}[wide, labelwidth=!, labelindent=0pt]
\item[\textbf{Step 1}] \noindent Initialize:
     \begin{itemize}
     \item Following Algorithm 4 in~\cite{stop_crit_2016}, estimate the number of recursive expansion iterations $n_{\text{max}}$, determine the acceleration parameters $\alpha_i$ and the polynomial sequence $p_i$, $i=1, 2,\ldots, n_{\text{max}}$, where $p_i = 1$ if $f_i(x) = ((1-\alpha_i) + \alpha_ix)^2$ and $p_i = 0$ if $f_i(x) = 2\alpha_ix-(\alpha_ix)^2$. Moreover, select the iteration $n_{\text{min}}$
      when acceleration should be switched off, i.~e.~$\alpha_i = 1$ for $i\geq n_{\text{min}}$. 
     \item Select truncation threshold values $\tau_i$, $i=0, 1,\ldots, n_{\text{max}}$ defined by~\eqref{eq:tau_trunc}.
     \item Rescale the matrix $F$ such that all its eigenvalues lie in the interval $[0,\ 1]$ in reverse order, see Algorithm~\ref{alg:init}. Here $\lambda_{\text{max}} \geq \lambda_N$ and $\lambda_{\text{min}} \leq \lambda_1$ are obtained using Gershgorin circles theorem. 
     \begin{algorithm}   
       \begin{algorithmic}[1]
         \caption{Initialization}
         \State $X_0 = \frac{\lambda_{\text{max}} I-F}{\lambda_{\text{max}}-\lambda_{\text{min}}}$
         \State $\wt{X}_0 = X_0+E_0$, $\norm{E_0}_M \leq \tau_0$  \Comment{Truncation}
         \State $e_0 = \norm{\wt{X}_0-\wt{X}_0^2}_F$
         \State $t_0 = \tr{\wt{X}_0-\wt{X}_0^2}$
         \label{alg:init}
         \end{algorithmic}
     \end{algorithm}
     \end{itemize}
  \item[\textbf{Step 2}] Recursively apply polynomials $p_i$ until the stopping criterion is satisfied, see Algorithm~\ref{alg:recexp}. The constant $C$ in line~\ref{line:alg_sp2acc} of Algorithm~\ref{alg:recexp} is used in the stopping criterion for the SP2 recursive expansion, and it is derived in~\cite{stop_crit_2016}.
  \begin{algorithm}   
    \begin{algorithmic}[1]
      \caption{The SP2ACC recursive expansion} 
      \State \textbf{Input:} $X_0$; $\alpha_i$, $p_i$, $\tau_i$, $i=1,\ldots,n_{\text{max}}$; $n_{\text{min}}$
      \State $C=6.8872$ \label{line:alg_sp2acc}\Comment{Needed for the stopping criterion}
      \For {$i = 1,2,\dots, n_{\text{max}}$} 
        \If {$p_i=1$}
        \State $X_{i} = ((1-\alpha_i)I + \alpha_i\wt{X}_{i-1})^2$
        \Else
        \State $X_{i} = 2\alpha_i\wt{X}_{i-1}-(\alpha_i\wt{X}_{i-1})^2$
        \EndIf
        \State $\wt{X}_i = X_i+E_i$, $\norm{E_i}_M \leq \tau_i$ \Comment{Truncation}
        \State $e_i = \norm{\wt{X}_i-\wt{X}_i^2}_F$
        \State $t_i = \tr{\wt{X}_i-\wt{X}_i^2}$
        \If {$i \geq n_{\text{min}}$ and $p_i \neq p_{i-1}$ and $e_i >Ce_{i-2}^2$}
            \State $n=i$
            \State \textbf{break}
        \EndIf
      \EndFor
      \State \textbf{Output:} $X_n$; $e_i$, $t_i$, $i=1,\ldots,n$
      \label{alg:recexp}
    \end{algorithmic}
    \label{alg:general_rec_exp_alg}
  \end{algorithm}
  
  \item[\textbf{Step 3}] 
  \begin{sloppypar}
    Compute homo and lumo eigenvalue estimates using Frobenius norms $e_i$ and traces $t_i$ following Algorithm 3 in~\cite{interior_eigenvalues_2014}. These estimates are then propagated to the next SCF cycle.
  \end{sloppypar}
\end{enumerate}

In Algorithms~\ref{alg:init} and ~\ref{alg:recexp} the truncation is written as an explicit addition of some error matrix $E_i$ in each iteration. In this work, we use truncation based on the Frob-Inf and Frobenius norms. The SP2 recursive expansion can be obtained using the algorithm outlined above with the acceleration parameters $\alpha_i$ set to 1 for all $i\geq 1$, and $n_{\text{min}} = 1$.

\section{Implementation details}

Chunks and Tasks~\cite{chunks-and-tasks} is a task-based programming model, where the runtime system is responsible for scheduling tasks and distributing the data into the physical resources. The idea is that the user splits the work into smaller tasks and the data into smaller chunks, and registers both to the library. As soon as chunks are registered to the library, they are read-only. Upon registration, chunks and tasks are assigned a unique identifier, which is chosen by the library. Since each task has input chunks which has to be registered to the library, it will never happen that some task requests non-existent chunks. Moreover, information about the location of the chunk can be stored in the chunk identifier, allowing efficient and easy data fetching. The CHT-MPI library~\cite{CHT-MPI_web} implements the Chunks and Tasks programming model. The task scheduler is based on randomized work stealing. As soon as one process runs out of work, it attempts to steal work from another process chosen at random. Tasks are scheduled for execution as soon as all input chunks are available, and each task can register new tasks and chunks.

Chunk objects may store different kinds of data, in particular they can
store chunk identifiers of other chunks, giving the possibility to create
hierarchical structures.
The Chunks and Tasks Matrix Library (CHTML) is a hierarchical matrix library written using the Chunks and Tasks programming model and implemented on top of the CHT-MPI library. In each level of the hierarchy, the matrix is divided into four blocks forming a quad-tree structure. The tree nodes on the lowest level of the hierarchy are called leaf nodes. Each non-leaf node stores chunk identifiers of corresponding non-zero submatrices. If the submatrix is zero, its chunk identifier is set to  \CPPcode{CHUNK\_ID\_NULL}. On the leaf level we use the block-sparse matrix representation where the leaf matrix is divided into small blocks and only elements of non-zero blocks are stored. Other representations of the leaf matrix are possible.

We implemented a set of classes, wrapping the CHTML matrix type and various matrix operations. As mentioned above, when a chunk is registered to the library, the user in return receives  a chunk identifier which can be used as an input to tasks. Inside the wrapper class we do not store any matrix elements, but only the identifier of a root chunk of a quad-tree representing the matrix.   The actual data may be distributed on different computational nodes by some previous calculation of the matrix. We provide alternative implementations of tasks for general and symmetric matrices. The symmetric matrix wrapper implementation utilizes matrix symmetry and assumes that only the upper triangle of the matrix is stored in memory.

\section{Numerical experiments}

In this section, we verify the efficiency of the density matrix construction implemented using CHTML in Ergo~\cite{ErgoSoftwareX}. We run our tests on the Rackham cluster at the UPPMAX computer center at Uppsala university, comprised of 486 nodes with two 10-core Intel Xeon V4 CPU's running at 2.2 GHz/core, each with at least 128 GB of RAM memory. We use the Chunks and Tasks matrix library linked to CHT-MPI 1.2~\cite{CHT-MPI_web} compiled with GCC 8.2.0 and Open MPI 3.1.3. The performance of leaf matrix operations is improved by linking to the optimized BLAS library OpenBlas 0.2.19 (single threaded).
We set maximum leaf matrix size to 4096 and small block size to 32. In the current implementation, the block size in the Frob-Inf norm is equal to the maximum leaf matrix size, i.e.~4096.  All timing results represent elapsed wall time.

To obtain the homo-lumo eigenvalue estimates we run the recursive expansion in each test twice, and measure time and communication cost for the second run. In the first run we do not have any assumption about the eigenvalues. Thus, we use the trace-correcting recursive expansion scheme~\cite{Nikl2002} with ad hoc truncation tolerances. At the end of the recursive expansion we obtain homo-lumo eigenvalue estimates and use them in the second run as described in section~\ref{sec:dens_constr}.

\subsection{Weak scaling tests on block-diagonal matrices}

\begin{figure}
  \begin{subfigure}[b]{.49\linewidth}
    \centering \includegraphics[width=\textwidth]{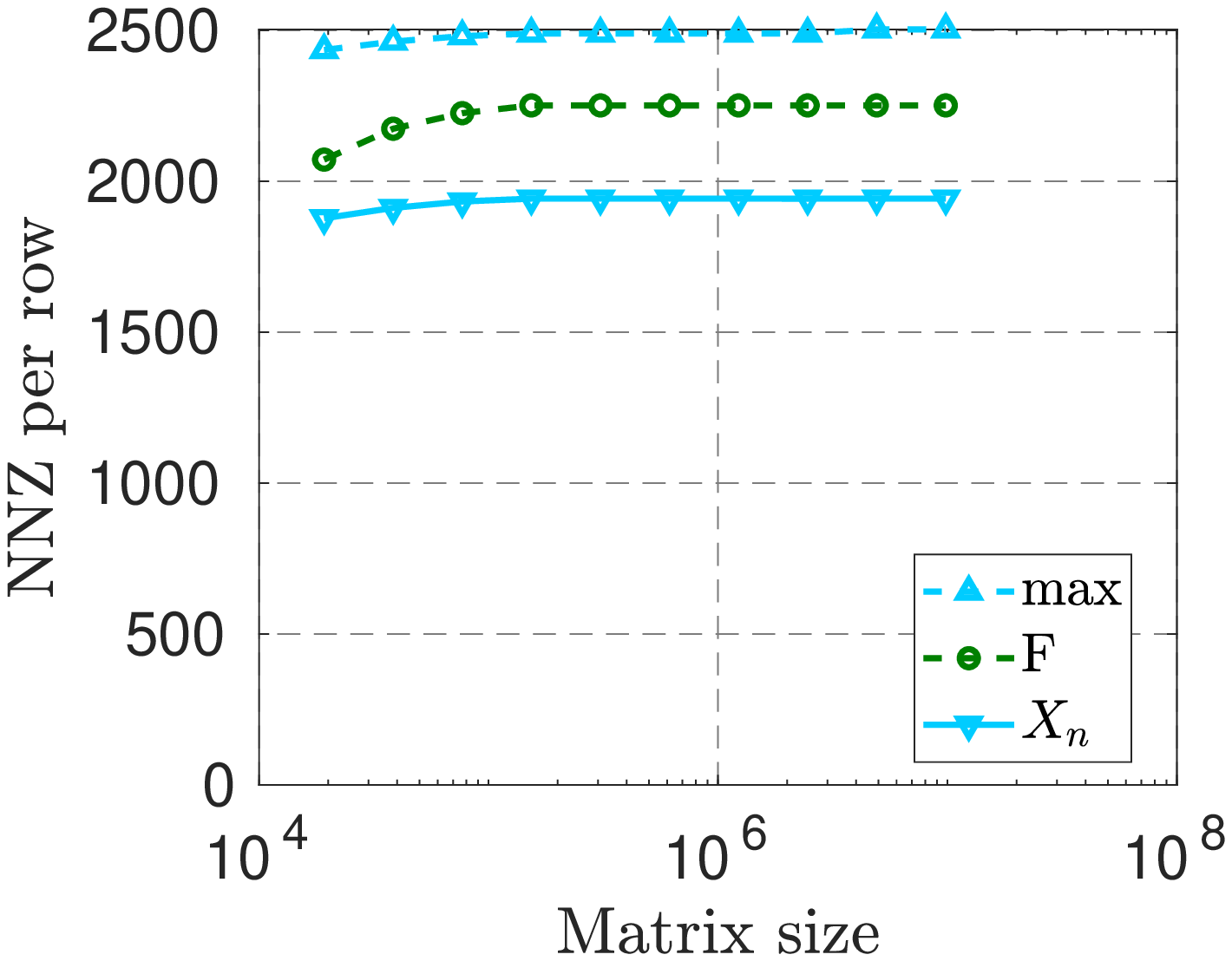}
    \caption{Number of non-zeros per row}\label{fig:gluala_nnz_per_row}
  \end{subfigure}
  \begin{subfigure}[b]{.49\linewidth}
    \centering \includegraphics[width=\textwidth]{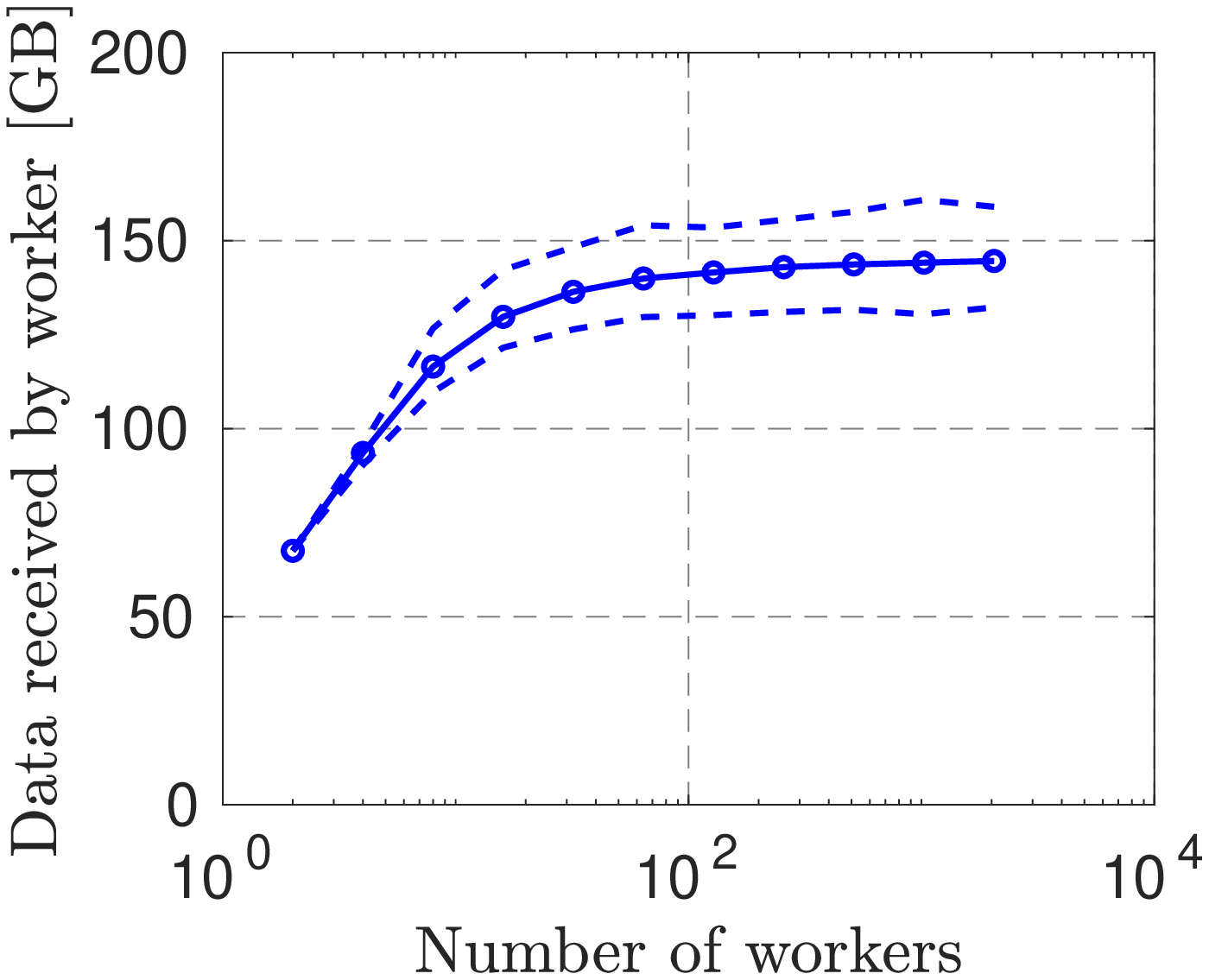}
    \caption{Data received by each worker}\label{fig:gluala_comm}
  \end{subfigure}
  \caption{Performance of the SP2ACC recursive expansion for the block-diagonal matrices, see text for construction details. Truncation is based on the Frob-Inf norm. Left panel: number of non-zero elements per row in the Fock matrix ($F$), in the obtained density matrix approximations ($X_n$), and the maximum number of non-zeros per row in all intermediate matrices $X_i$ throughout all recursive expansion iterations (max). Right panel: the solid line presents the average communication cost per worker process over all workers and all recursive expansion iterations. Maximum and minimum values over all workers and all iterations are plotted using dashed lines.}\label{fig:gluala}
\end{figure}

In a weak scaling test the amount of work per worker process is kept fixed. In order to achieve this, the number of non-zero elements per row in all (Fock, intermediate and final density) matrices should be kept constant with increasing system size. Moreover, to get the same number of recursive expansion iterations, the homo-lumo gap should be kept constant too.
We construct block-diagonal matrices of increasing size with the desired properties mentioned above. In the created matrices we repeat the same test matrix along the diagonal. The test matrix is a Fock matrix of size 19204  obtained in the last SCF cycle of Hartree--Fock calculations using the standard Gaussian basis set 3-21G on a glutamic acid and alanine (Glu-Ala) helix with 3330 atoms. The estimated number of SP2ACC recursive expansion iterations is 24 for all cases and the actual number of iterations is 18 for the smallest matrix size and 20 for all other matrices. The maximum allowed error in the occupied subspace is set to $0.001$. Truncation is based on the Frob-Inf norm. In Figure~\ref{fig:gluala_nnz_per_row} we show that the number of non-zero elements per row is constant for large enough matrices.
In our tests we increase the number of blocks $K$ along the diagonal proportionally to the number of worker processes $N_p$ such that $N_p = 2K$. Since we are interested in the communication cost per worker process and not in the actual timings, we put one worker process per core and use only one worker thread per worker. We do not use the chunk cache, \ie it is set to 0 GB. 
In Figure~\ref{fig:gluala_comm} we demonstrate that the amount of data received per worker process tends to a constant with increasing number of workers. We note that these results are in agreement with the results obtained in~\cite{LocalityAwareRubensson2016}, where only the sparse matrix-matrix multiplication is analyzed.

We note that this is an illustrative example. The ideal distribution of work and data over processes leads to separate computation of each diagonal block, without any communication of matrix elements between blocks. However, if the random permutation is used to achieve load balance, such trivial non-zero matrix structure cannot be exploited, resulting in unnecessary communication of matrix elements. In our approach, the load balance is achieved dynamically, and we are able to utilize the non-zero matrix structure reducing the communication cost.

\subsection{Calculations on water clusters}

In this section we present results of density matrix calculations for water clusters. Water cluster geometries are generated from a large molecular dynamics simulation of bulk water at standard temperature and pressure by including all water molecules within spheres of varying radii. We perform 
calculations using our test implementation of the Hartree--Fock method parallelized for distributed-memory. We use the Gaussian basis set 3-21G. Initial guess density matrices are obtained from calculations with a smaller basis set STO-2G. We perform 5 SCF cycles and save Fock matrices in the orthogonal basis in the last cycle. The recursive expansion is applied on the saved Fock matrices. 
The chunk cache size is set to 8 GB per worker. Two workers are placed on each node (using \texttt{--bind-to socket -map-by socket} Open MPI flags) and each worker has 9 worker threads. 
The maximum allowed error in the occupied subspace is set to $0.001$. We note that the homo-lumo gap does not change significantly with increasing problem size. The estimated number of iterations in SP2 and SP2ACC recursive expansions is 35-37 and 22-23, respectively. The obtained number of recursive expansion iterations in SP2 and SP2ACC recursive expansions is 29 and 16-17, respectively.

\begin{figure}[!ht]
  \begin{subfigure}[b]{.49\linewidth}
    \centering \includegraphics[width=\textwidth]
        {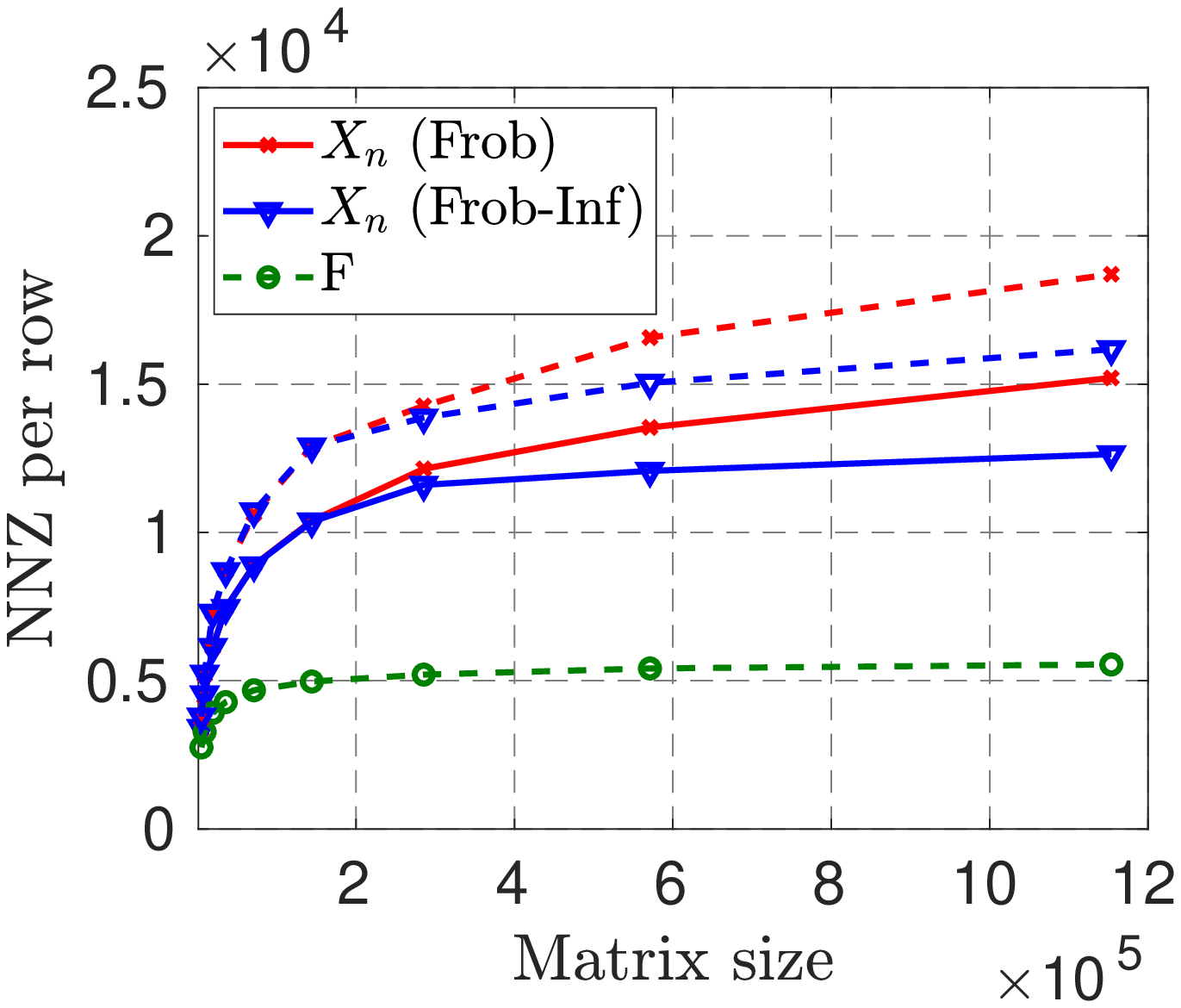}
    \caption{SP2}\label{fig:nnz_per_row_water_sp2}
  \end{subfigure}
  \begin{subfigure}[b]{.49\linewidth}
    \centering \includegraphics[width=\textwidth]
        {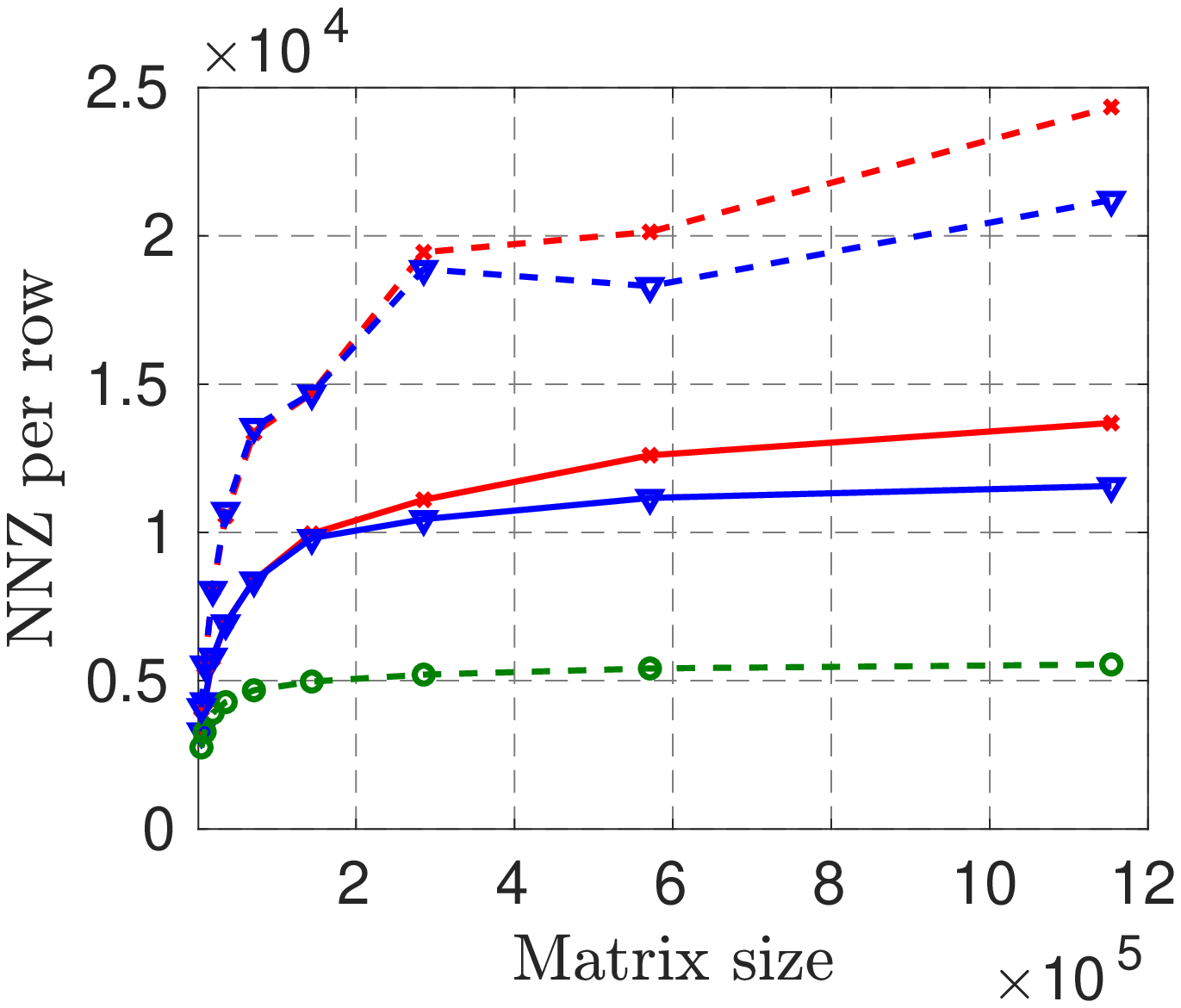}
    \caption{SP2ACC}\label{fig:nnz_per_row_water_sp2acc}
  \end{subfigure}
  \caption{Number of non-zero elements per row in Fock matrices ($F$), in the obtained density matrix approximations ($X_n$), and the maximum number of non-zeros per row in all intermediate matrices $X_i$ throughout all recursive expansion iterations (plotted using dashed lines). Fock matrices are obtained from HF/3-21G calculations for water clusters of increasing size. Lines with marker ``$\times$'' correspond to calculations with truncation based on the Frobenius norm, and lines with marker ``$\triangledown$'' correspond to calculations with truncation based on the Frob-Inf norm.
  }
  \label{fig:nnz_per_row_water}
\end{figure}

In our tests the number of nodes is increasing proportionally to the matrix size. Number of non-zeros per row in the Fock matrices, obtained density matrices and the maximum number of non-zeros per row between all intermediate matrices $X_i$ are presented in Figure~\ref{fig:nnz_per_row_water} for both SP2 and SP2ACC recursive expansions. 

\begin{figure}[!ht]
  \begin{subfigure}[b]{.49\linewidth}
    \centering \includegraphics[width=\textwidth]
        {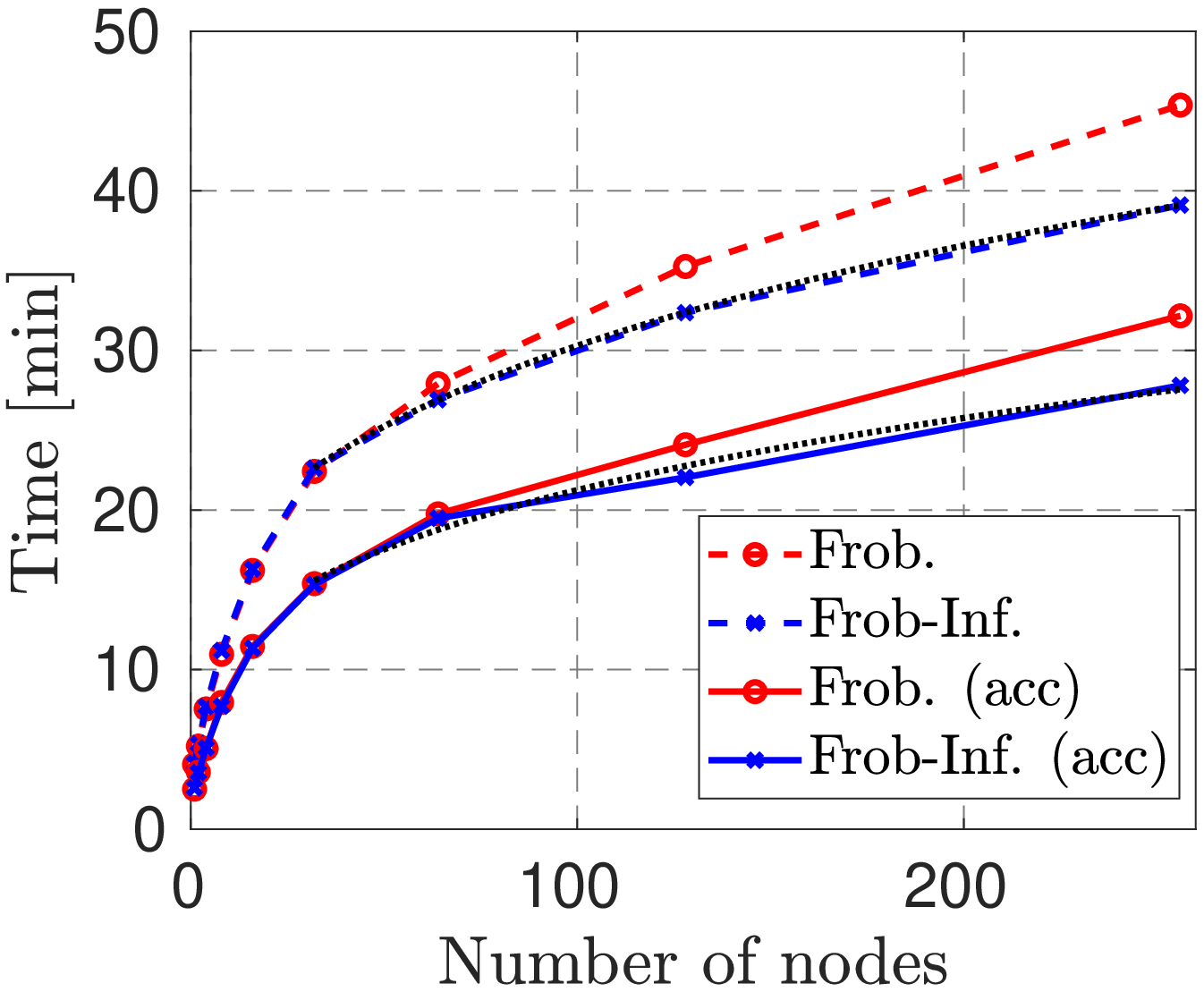}
    \caption{Execution time}\label{fig:linear_scaling_water_norms_time}
  \end{subfigure}
  \begin{subfigure}[b]{.49\linewidth}
    \centering \includegraphics[width=\textwidth]
        {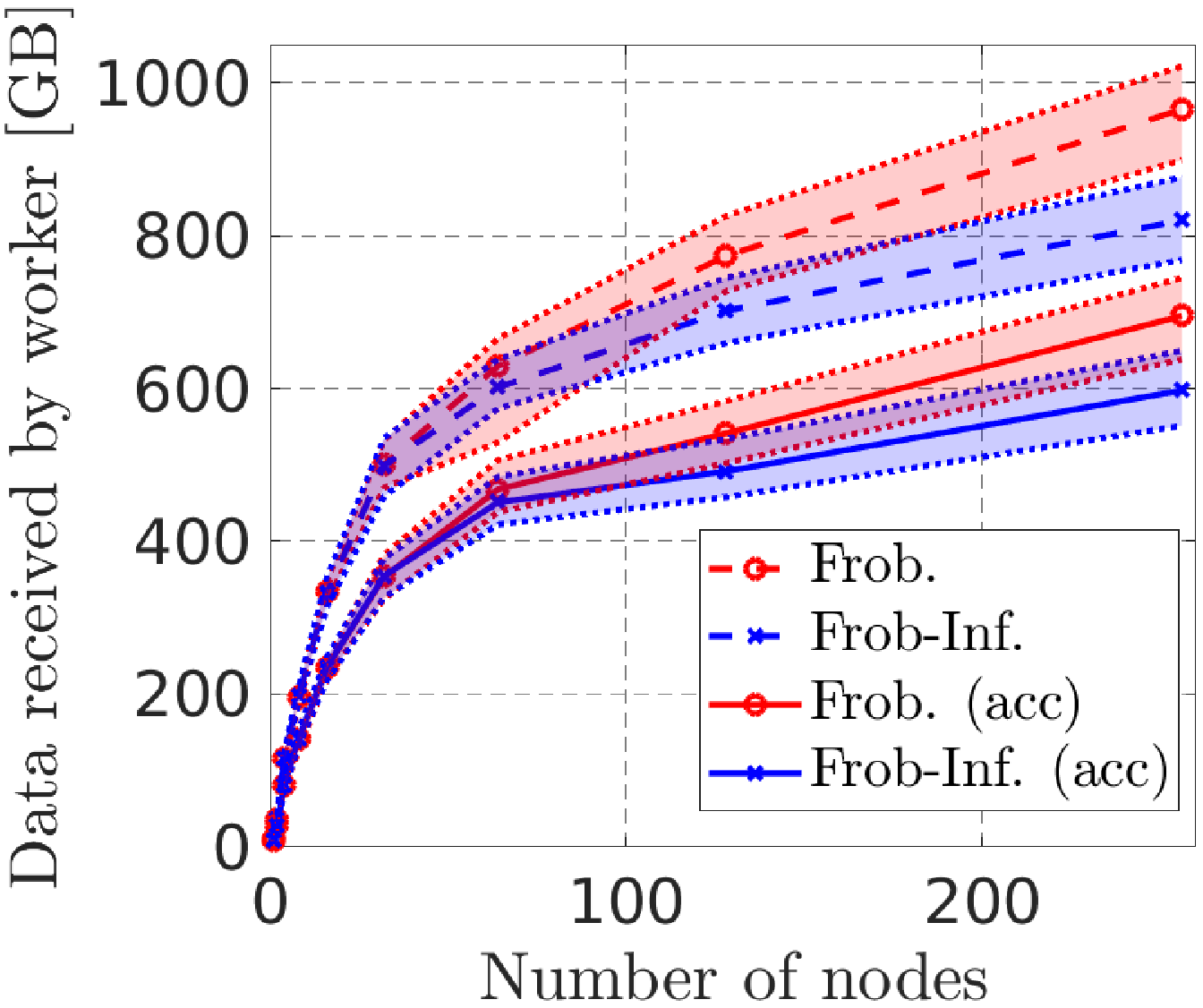}
    \caption{Communication cost}\label{fig:linear_scaling_water_norms_comm}
  \end{subfigure}
  \caption{Performance comparison of the SP2 (dashed lines) and SP2ACC (``acc'', solid lines) recursive expansions with truncation based on the Frobenius (``Frob.'', ``$\circ$'') and Frob-Inf (``$\times$'') norms. Fock matrices are obtained from HF/3-21G calculations for water clusters of increasing size. In the left panel dotted black lines show the least square fit of data for the Frob-Inf norm by a function $c_0 + c_1 \log{(N_p)}+c_2\left(\log{(N_p)}\right)^2$. 
  In the right panel we show the average, maximum and minimum communication cost over all workers and all iterations. Maximum and minimum values are plotted using dotted lines and the area in between is shaded. }\label{fig:linear_scaling_water_norms}
\end{figure}

In Figure~\ref{fig:linear_scaling_water_norms} we compare the execution time and the communication cost of calculations where truncation is based on the  Frobenius and Frob-Inf norms. We use the wrapper class representing a general matrix.
In~\cite{LocalityAwareRubensson2016}, it is shown that the execution time of sparse matrix-matrix multiplication implemented in CHTML scales as $\OO((\log{(N_p)})^2)$ where $N_p$ is a number of worker processes.
In Figure~\ref{fig:linear_scaling_water_norms_time}, data indicate that the same result is obtained for the whole recursive expansion.

\begin{figure}[!ht]
  \begin{subfigure}[b]{.49\linewidth}
    \centering \includegraphics[width=\textwidth]
        {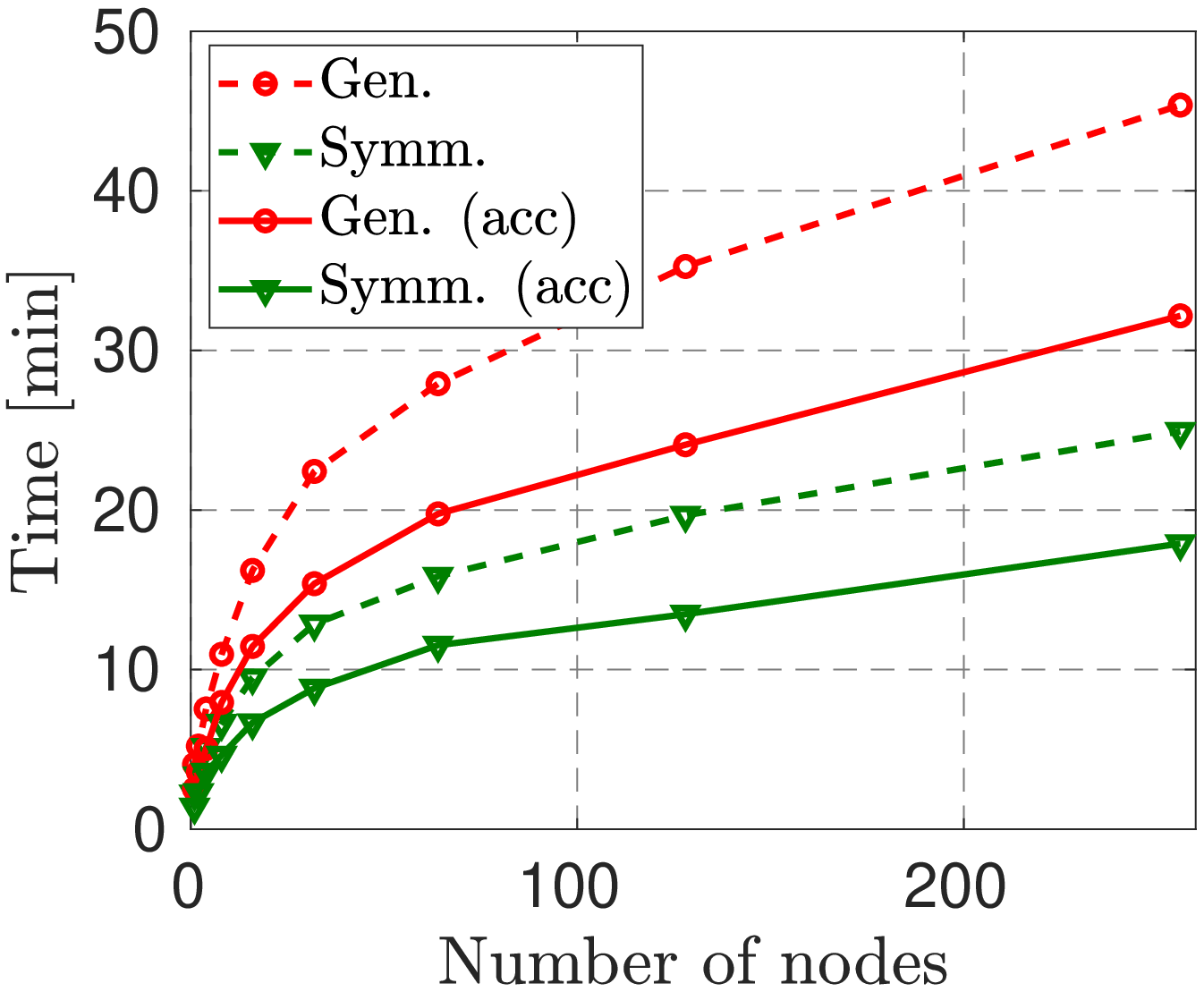}
    \caption{Execution time}\label{fig:linear_scaling_water_gen_symm_time}
  \end{subfigure}
  \begin{subfigure}[b]{.49\linewidth}
    \centering \includegraphics[width=\textwidth]
        {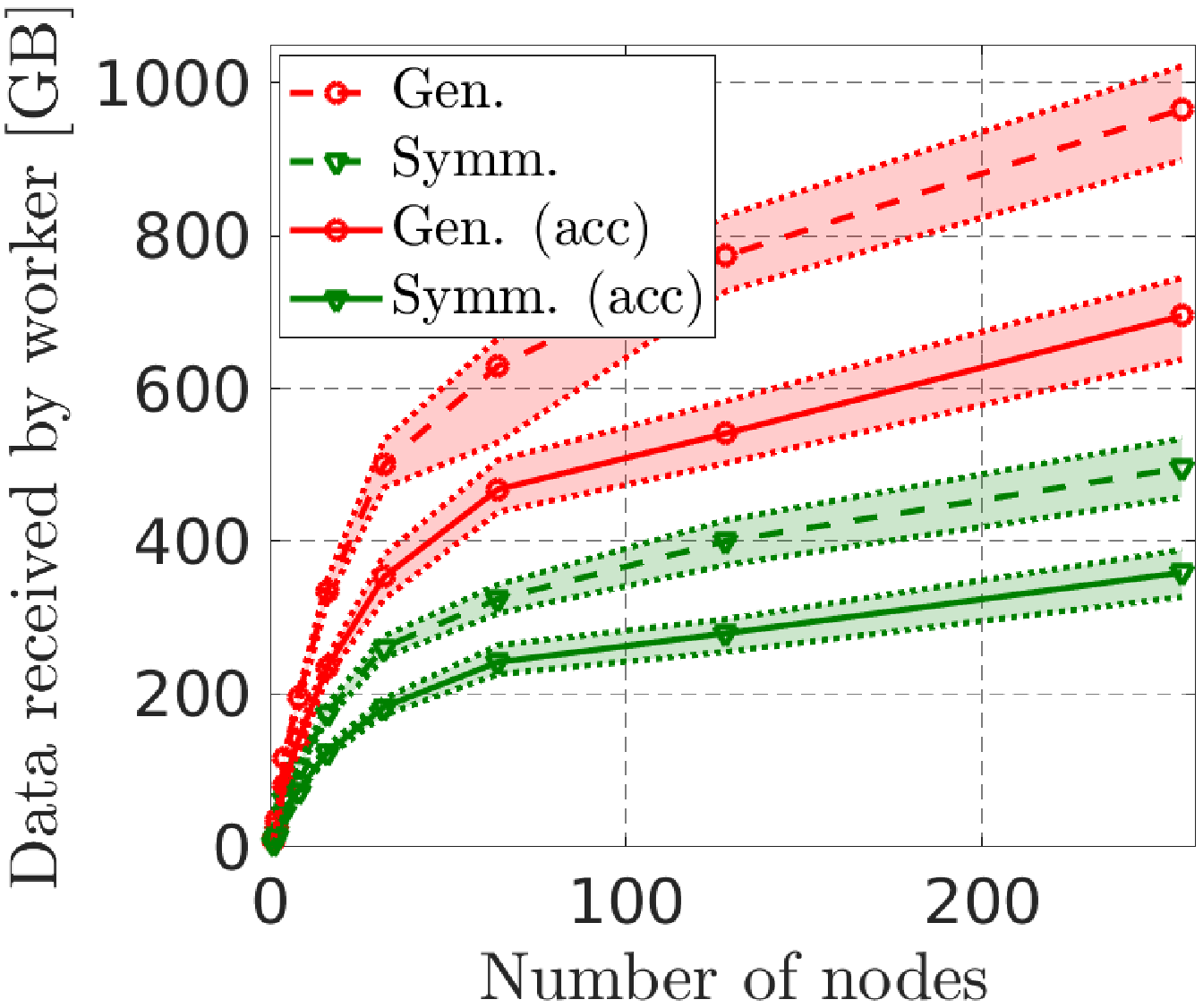}
    \caption{Communication cost}\label{fig:linear_scaling_water_gen_symm_comm}
  \end{subfigure}
  \caption{
  Performance comparison of the SP2 (dashed lines) and SP2ACC (``acc'', solid lines) recursive expansions implemented using the CHTML wrappers for general (``$\circ$'') and symmetric (``$\triangledown$'') matrices.
  Truncation is based on the Frobenius norm. Fock matrices are obtained from HF/3-21G calculations for water clusters of increasing size. 
    In the right panel we show the average, maximum and minimum communication cost over all workers and all iterations. Maximum and minimum values are plotted using dotted lines and the area in between is shaded.
  }\label{fig:linear_scaling_water_gen_symm}
\end{figure}

We perform recursive expansions using the CHTML wrapper for general matrices where matrix symmetry is ignored, and the wrapper for symmetric matrices where only the upper triangle of the matrix is stored and manipulated. Truncation based on the Frobenius norm is used in both cases. The total execution time and the amount of data received per worker process are given in Figure~\ref{fig:linear_scaling_water_gen_symm}. The figure shows that CHTML with the block-sparse leaf matrix is able to utilize matrix symmetry, providing a reduction of the total execution time and communication cost by almost a factor of two. 
%
%


\section{Concluding remarks}

In this work, the density matrix is constructed using the SP2ACC recursive polynomial expansion, which scales linearly with system size for sufficiently sparse matrices. The matrix sparsity is enforced by removal of small matrix elements. The combination of the error control scheme~\cite{ErrorControl} and parameterless stopping criteria~\cite{stop_crit_2016} enables us to use only one user-defined tolerance for the density matrix computation: the maximum allowed error in the occupied invariant subspace. Truncation tolerances are  chosen dynamically at the beginning of the recursive expansion to ensure the desired accuracy of the density matrix approximation.

We study the execution time and communication cost of the recursive polynomial expansion implemented using the Chunks and Tasks matrix library (CHTML)~\cite{LocalityAwareRubensson2016,chunks-and-tasks}. Matrices are expressed as quad-trees of chunk objects, where the matrix elements are stored on the lowest level using a block-sparse matrix format. The parallelism is then expressed using tasks operating on the matrix hierarchies. The sparse matrix-matrix multiplication is an important building block of the recursive expansion. Here we use the locality-aware sparse matrix-matrix multiplication implemented in CHTML~\cite{LocalityAwareRubensson2016}. 
We demonstrate that, if the number of recursive expansion iterations and the number of non-zero elements per row in all matrices is bounded with increase of the system size, the communication cost is asymptotically constant. Moreover, we show that the quad-tree matrix representation in CHTML allows for efficient symmetry utilization.

To the best of our knowledge this is the first implementation of the density matrix construction with rigorous error control of the density matrix approximation and bounded communication cost in a weak scaling limit. The current study focuses on the computation of the density matrix from a given Fock/Kohn-Sham matrix. Future work includes parallelization using Chunks and Tasks of the SCF procedure implemented in the Ergo code~\cite{ErgoSoftwareX}.

\section*{Acknowledgments}

This work was supported by the Swedish strategic research programme eSSENCE.
Computational resources were provided by the Swedish National
Infrastructure for Computing (SNIC) through Uppsala Multidisciplinary Center for Advanced Computational Science (UPPMAX) in Uppsala, Sweden.

\bibliography{biblio} \bibliographystyle{siam} 

\begin{thebibliography}{10}

\bibitem{azad2016exploiting}
{\sc A.~Azad, G.~Ballard, A.~Bulu{\c{c}}, J.~Demmel, L.~Grigori, O.~Schwartz,
  S.~Toledo, and S.~Williams}, {\em Exploiting multiple levels of parallelism
  in sparse matrix-matrix multiplication}, SIAM J. Sci. Comput., 38 (2016),
  pp.~C624--C651.

\bibitem{ballard2013communication}
{\sc G.~Ballard, A.~Bulu{\c{c}}, J.~Demmel, L.~Grigori, B.~Lipshitz,
  O.~Schwartz, and S.~Toledo}, {\em Communication optimal parallel
  multiplication of sparse random matrices}, in Proceedings of the twenty-fifth
  annual ACM symposium on Parallelism in algorithms and architectures, ACM,
  2013, pp.~222--231.

\bibitem{benzi_decay}
{\sc M.~Benzi, P.~Boito, and N.~Razouk}, {\em Decay properties of spectral
  projectors with applications to electronic structure}, SIAM Rev., 55 (2013),
  pp.~3--64.

\bibitem{bulucc2012parallel}
{\sc A.~Bulu{\c{c}} and J.~R. Gilbert}, {\em Parallel sparse matrix-matrix
  multiplication and indexing: {I}mplementation and experiments}, SIAM J. Sci.
  Comput., 34 (2012), pp.~C170--C191.

\bibitem{LATTE-jcp-2012}
{\sc M.~J. Cawkwell and A.~M.~N. Niklasson}, {\em Energy conserving, linear
  scaling {Born-Oppenheimer} molecular dynamics}, J. Chem. Phys., 137 (2012),
  p.~134105.

\bibitem{dawson2018massively}
{\sc W.~Dawson and T.~Nakajima}, {\em Massively parallel sparse matrix function
  calculations with {NTP}oly}, Comput. Phys. Commun., 225 (2018), pp.~154--165.

\bibitem{hohen}
{\sc P.~Hohenberg and W.~Kohn}, {\em Inhomogeneous electron gas}, Phys. Rev.,
  136 (1964), pp.~B864--B871.

\bibitem{horn2012matrix}
{\sc R.~A. Horn and C.~R. Johnson}, {\em Matrix analysis}, Cambridge University
  Press, New York, USA, 2012.

\bibitem{KohnSham65}
{\sc W.~Kohn and L.~J. Sham}, {\em Self-consistent equations including exchange
  and correlation effects}, Phys. Rev., 140 (1965), p.~1133.

\bibitem{stop_crit_2016}
{\sc A.~Kruchinina, E.~Rudberg, and E.~H. Rubensson}, {\em Parameterless
  stopping criteria for recursive density matrix expansions}, J. Chem. Theory
  Comput., 12 (2016), pp.~5788--5802.

\bibitem{lazzaro2017increasing}
{\sc A.~Lazzaro, J.~VandeVondele, J.~Hutter, and O.~Sch{\"u}tt}, {\em
  Increasing the efficiency of sparse matrix-matrix multiplication with a 2.5
  {D} algorithm and one-sided {MPI}}, in Proceedings of the Platform for
  Advanced Scientific Computing Conference, ACM, 2017, p.~3.

\bibitem{Nikl2002}
{\sc A.~M.~N. Niklasson}, {\em Expansion algorithm for the density matrix},
  Phys. Rev. B, 66 (2002), p.~155115.

\bibitem{honpas}
{\sc X.~Qin, H.~Shang, H.~Xiang, Z.~Li, and J.~Yang}, {\em {HONPAS}: A linear
  scaling open-source solution for large system simulations}, Int. J. Quantum
  Chem., 115 (2015), pp.~647--655.

\bibitem{Roothaan}
{\sc C.~C.~J. Roothaan}, {\em New developments in molecular orbital theory},
  Rev. Mod. Phys., 23 (1951), pp.~69--89.

\bibitem{Rub2011}
{\sc E.~H. Rubensson}, {\em Nonmonotonic recursive polynomial expansions for
  linear scaling calculation of the density matrix}, J. Chem. Theory Comput., 7
  (2011), pp.~1233--1236.

\bibitem{localized_inverse_factorization}
{\sc E.~H. Rubensson, A.~G. Artemov, A.~Kruchinina, and E.~Rudberg}, {\em
  Localized inverse factorization}, arXiv:1812.04919,  (2019).

\bibitem{interior_eigenvalues_2014}
{\sc E.~H. Rubensson and A.~M.~N. Niklasson}, {\em Interior eigenvalues from
  density matrix expansions in quantum mechanical molecular dynamics}, SIAM J.
  Sci. Comput., 36 (2014), pp.~B147--B170.

\bibitem{CHT-MPI_web}
{\sc E.~H. Rubensson and E.~Rudberg}, {\em CHT-MPI}, Available at
  \href{http://chunks-and-tasks.org}{http://chunks-and-tasks.org} (Accessed
  September 2019).

\bibitem{mixedNormTrunc}
\leavevmode\vrule height 2pt depth -1.6pt width 23pt, {\em Bringing about
  matrix sparsity in linear scaling electronic structure calculations}, J.
  Comput. Chem., 32 (2011), pp.~1411--1423.

\bibitem{chunks-and-tasks}
\leavevmode\vrule height 2pt depth -1.6pt width 23pt, {\em Chunks and {T}asks:
  {A} programming model for parallelization of dynamic algorithms}, Parallel
  Comput., 40 (2014), pp.~328--343.

\bibitem{LocalityAwareRubensson2016}
\leavevmode\vrule height 2pt depth -1.6pt width 23pt, {\em Locality-aware
  parallel block-sparse matrix-matrix multiplication using the {C}hunks and
  {T}asks programming model}, Parallel Comput., 57 (2016), pp.~87--106.

\bibitem{ErrorControl}
{\sc E.~H. Rubensson, E.~Rudberg, and P.~Sa{\l}ek}, {\em Density matrix
  purification with rigorous error control}, J. Chem. Phys., 128 (2008),
  p.~074106.

\bibitem{ergo_web}
{\sc E.~Rudberg, E.~H. Rubensson, P.~Sa{\l}ek, and A.~Kruchinina}, {\em Ergo
  (Version~3.8). Available at \url{http://www.ergoscf.org} (Accessed September
  2019)}.

\bibitem{ErgoSoftwareX}
\leavevmode\vrule height 2pt depth -1.6pt width 23pt, {\em Ergo: {A}n
  open-source program for linear-scaling electronic structure calculations},
  SoftwareX, 7 (2018), pp.~107--111.

\bibitem{VandeVondele_2012}
{\sc J.~VandeVondele, U.~Bor\v{s}tnik, and J.~Hutter}, {\em Linear scaling
  self-consistent field calculations with millions of atoms in the condensed
  phase}, J. Chem. Theory Comput., 8 (2012), pp.~3565--3573.

\end{thebibliography}

\end{document}